\documentclass[]{aa}
\usepackage[varg]{txfonts}
\pdfoutput=1
\bibpunct{(}{)}{;}{a}{}{,} 
\usepackage{amssymb}
\usepackage{amsmath}
\usepackage{color}

\begin{document}

\title{Mass-metallicity relation of dwarf galaxies and its dependency on time: clues from resolved systems and comparison with massive galaxies}
\titlerunning{Mass-metallicity relation of dwarf galaxies and its dependency on time}

\author{S.~L. Hidalgo\inst{1,2}}
\institute{Instituto de Astrof\'\i sica de Canarias. V\'\i a L\'actea s/n. E38200 - La Laguna, Tenerife, Canary Islands, Spain. \email{shidalgo@iac.es}
\and 
Department of Astrophysics, University of La Laguna. V\'\i a L\'actea s/n. E38200 - La Laguna, Tenerife, Canary Islands, Spain}

\abstract{}{We present a new approach to study the mass-metallicity relation and its dependency on time.}{We used the star formation history (SFH)\ derived from color-magnitude diagram fitting techniques of a sample of Local Group (LG) dwarfs to obtain stellar masses, metallicities, and star-formation rates (SFR) to analyze the mass-metallicity relation as a function of the ages of their stellar populations. The accurate SFHs allow a time resolution of about 2 Gyr at the oldest ages for a total redshift range of $0\lesssim z\lesssim 3$.}{The mass-metallicity relation retrieved for the sample of LG dwarfs was compared with a large dataset of literature data obtained in a wide redshift range. Neither
of the two independent datasets shows a clear evolution of the mass-metallicity relation slope with redshift. However, when the star-formation rate is added as an additional parameter in the relation, it shows a dependence on the redshift in the sense that the coefficient of the mass decreases with increasing redshift, while the coefficient for the SFR is almost constant with time. This result suggests an increasing contribution with time of the galaxy stellar mass to the metalliticy of the stars that
formed most recently, but it also shows that the SFR can play a fundamental role in shaping the mass-metallicity relation.}{}

\maketitle

\section{INTRODUCTION}\label{secint}

Galaxies show a tight luminosity-metallicity relation in which those with higher luminosities also have higher metallicities. This relation has been attributed to a more generalized relation, the mass-metallicity relation \citep{skillmanetal1989,tremontietal2004,kirbyetal2013}.
The mass-metallicity relation, $Z(M_\ast)$, has been observed in the nearest galaxies, most of them located in the Local Group, using gas-phase metallicities (\citealp{skillmanetal1989}; \citealp{vanzeeetal1997,leeetal2006}), but also using spectroscopic metallicities of individual stars \citep{kirbyetal2013}. Gas-phase metallicites are usually obtained from HII regions or planetary nebula, while red giant branch stars (RGB) are used to obtain spectroscopic metallicities of individual stars. For a significant number of samples, the mean metallicities obtained from HII regions are an estimate of the current mean metallicity of a galaxy (i.e., a spatial average), while the RGB metallicities represent the averaged metallicity that is locked up in stars older than ~2 Gyr (i.e., a time average). Despite this significant difference, the slopes of the $Z(M_\ast)$ relation based on the two different metallicity estimates are in fine agreement \citep{kirbyetal2013}.

The study of the zero-point and slope of $Z(M_\ast)$ and its dependency on time may shed light on how galaxies form, on the
galaxy assembling history, the rate of collecting fresh gas, or the dominant chemical evolution model in each phase \citep{savaglioetal2005,maiolinoetal2008,molleretal2013}. Great effort has been made to study this dependency by increasing the number and redshift range of galaxies. Starting with the early works by \citet{richer&mccall1995}, \citet{vanzeeetal1997}, and \citet{skillmanetal1989}, the study of $Z(M_\ast)$ has been carried out in different surveys in the last years: \citet[DGSS]{kobulnickyetal2003}; \citet[GOODS-N]{kobulnicky&kewley2004}; \citet[SDSS]{tremontietal2004,mannuccietal2010,zahidetal2012}; \citet[GDDS]{savaglioetal2005}; \citet[AMAZE-I]{maiolinoetal2008}, and \citet[GAMA]{fosteretal2012} are some examples.

In all these works, oxygen abundances show a direct correlation with both the absolute magnitude of the galaxies in B band, $\rm M_B$, and the galaxy stellar masses, $\rm M_\ast$ , although the correlation flattens at $\rm M_\ast\ga 10^{10.5}~M_\odot$  \citep{tremontietal2004,leeetal2006,kewley&ellison2008,mannuccietal2010,yatesetal2012}. Using the results provided by this type of surveys, \citet{carrollo&lilly2001,crescietal2012,lamareilleetal2006} reported no significant evolution of $Z(M_\ast)$ with redshift. \citet{mannuccietal2010} and \citet{crescietal2010} reported no evolution of $Z(M_\ast)$ during 80\% and 50\% of cosmic history, respectively. Modest evolutionary effects have been found by \citet{lillyetal2003, kobulnicky&kewley2004} in the zero-point, or in the shape by \citet{fosteretal2012}. On the other hand, several papers have found a clear evolution of the zero-point \citep{savaglioetal2005,maiolinoetal2008,molleretal2013} and in the slope of the mass-metallicity relation \citep{maiolinoetal2008,caluraetal2009,lara-lopezetal2010}. 

The discrepancy among these results may have different sources: uncertainties associated with the derivation of metallicities, magnitudes, and masses of galaxies at high redshift ($z\ga 2.5$), poor statistics (only galaxies with $z\la 0.5$ have been well sampled), or even a selection effect of the galaxies with redshift (e.g., higher star formation or different morphological types) \citep{mannuccietal2010}. Finally, \citet{kewley&ellison2008} have pointed out that the largest source of uncertainty is related to the dependency of the shape of $Z(M_\ast)$ on the metallicity callibrators and redshift.

\citet{kewley&ellison2008} provided compelling evidence about a difference up to $\rm \Delta[log (O/H)] = 0.7$~dex in the absolute metallicity scale with a substantial change in the shape of $Z(M_\ast)$.  \citet{kewley&ellison2008} showed that by selecting the correct metallicity conversions, the difference in the absolute metallicity scale can be reduced to $\sim 0.03$~dex. The transformation of the metallicity estimates to a homogeneous scale is then mandatory before performing any analysis of $Z(M_\ast)$  as a function of redshift.

In addition to all these considerations, a new approach to understand the nature of $Z(M_\ast)$ has been proposed in the last years: the observed $Z(M_\ast)$ could be due to a more general relation involving other parameters such as the star formation rate (SFR) \citep{yatesetal2012,mannuccietal2010,crescietal2012} or the HI content of the galaxies \citep{bothwelletal2013,lara-lopezetal2013}. \citet{ellisonetal2008}, \citet*{peeplesetal2008}, \citet{mannuccietal2010}, and  \citet{yatesetal2012} found that low-mass galaxies with higher SFR have lower metallicities. \citet{bothwelletal2013} found that HI-rich galaxies are more metal poor at a given stellar mass, pointing to a strong secondary dependency of the $Z(M_\ast)$ on HI mass. On the other hand, \citet{lara-lopezetal2013} found a correlation between $Z(M_\ast)$, the SFR, and the gas content of the galaxies: the low-mass galaxies with a large amount of gas have low metallicities, while those with small amounts of gas show high metallicities. 

We propose here a new approach in the study of $Z(M_\ast)$ and its dependency on time. We use the star formation history (SFH) obtained from CMD-fitting techniques to obtain ages, metallicities, and the stellar mass of a sample of galaxies, which is compared with gas-phase metallicities from the literature. For the present work, we adopt the very detailed and accurate SFHs obtained for a sample of isolated Local Group dwarfs in the framework of the local cosmology of isolated dwarfs (hereafter, LCID dwarfs) Leo-A, Phoenix, Cetus, Tucana, LGS-3, and IC1613 \citep{coleetal2007,hidalgoetal2009,monellietal2010a,monellietal2010b,hidalgoetal2011,skillmanetal2014}. These galaxies show evidence for a dependency of the metallicity on the cumulative mass function \citep{hidalgoetal2013}, pointing to a mass-metallicity relation. The use of the LCID dwarfs data allows us to investigate how the mass-metallicity relation changes with time in a long time interval. This information can be compared with suitable literature data, providing clues about the same relation at different redshifts. We wish to note that although the sample of LCID dwarfs is quite small when compared with the sample of galaxies studied in the surveys devoted to investigate the $Z(M_\ast)$ relation, the fact that all LCID dwarfs are analyzed with a homogeneous methodological approach makes them a very suitable dataset to study the $Z(M_\ast)$ relation and its possible time dependence. 

The organization of this paper is as follows: the mass-metallicity relation obtained using both gas-phase metallicities and spectroscopy from resolved stars is compared with the relation obtained from CMD-fitting techniques in Sect. \ref{secmm}. The mass-metallicity relation from gas-phase metallicities and its dependency on time is analyzed in Sect. \ref{secmmt}. A multivariate analysis of the mass-metalliticy relation is performed in Sect. \ref{secfmm}. Finally, the results are discussed and summarized in Sect. \ref{secdisc}.

For the conversion between redshift and the age of the stellar populations obtained from SFHs, we assume a flat Einstein-de Sitter universe with $H_0=71.9~\rm km~s^{-1}~Mpc^{-1} $ and $\Omega_m=0.258$, as derived from \citet{komatsuetal2009}.

\section{LOCAL MASS-METALLICTY RELATION}\label{secmm}

The SFH is a very convenient tool to study the evolution of galaxies since the onset of the star formation at the beginning of their evolution. We used the SFH data for LCID dwarfs in order to investigate the dependence on time of the mass-metallicity relation. These results are based on accurate SFHs derived from deep color-magnitude diagrams (CMDs) reaching the oldest main-sequence turn-offs. 

In short, the CMD-fitting technique compares the star distribution in an observed CMD with the distribution in a synthetic CMD that serves as a model. The synthetic CMD is built with no assumption about the SFR or the metallicity distribution. The synthetic CMD is divided into stellar populations (i.e., a number of stars within a small range of age and metallicity), and the distribution of these stellar populations in the synthetic CMD is compared with the distribution of stars in the observed CMD by sampling CMDs in boxes and counting stars in them. A merit function is used to select the combination of weights of each stellar population that best represents the observed CMD in terms of the number of stars in each box. The solution, the SFH, provides the rate at which stars are formed, $\psi(t)$, and the metallicity of the most recently formed stars, $Z(t)$, at each time snapshot, $t$, into which the stellar populations are divided.

The detailed method to obtain a SFH using a CMD can be found in \citet{aparicio&hidalgo2009} and \citet{hidalgoetal2011}. The results for the individual galaxies Leo-A, Phoenix, Cetus, Tucana, LGS-3, and IC1613 can be found in \citet{coleetal2007,hidalgoetal2009,monellietal2010a,monellietal2010b,hidalgoetal2011,skillmanetal2014}, respectively. Since the LCID dwarfs are not completely covered by the field of view in the observations, we used $\rm M_V$ given in \citet{mcconnachie2012} to scale $\rm M_\ast$ and $\psi(t)$ for each galaxy.

\begin{figure}
\centering
\includegraphics[width=8.4cm]{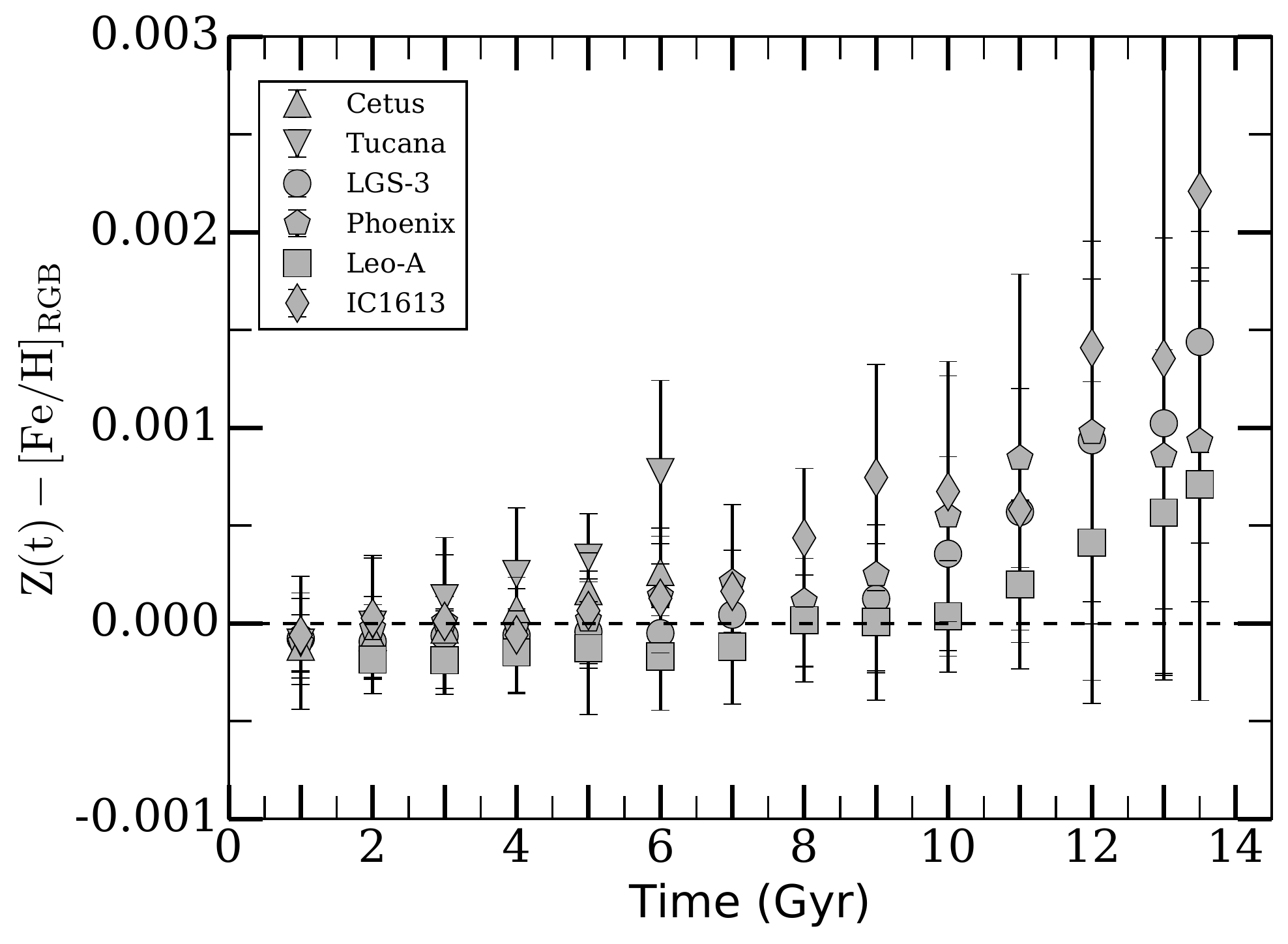}
\protect\caption[]{Difference between $\rm Z(t)$ and $\rm [Fe/H]_{RGB}$ as a function of time for the LCID sample. The horizontal dashed line shows $\rm Z(t) = [Fe/H]_{RGB}$.}
\label{f00}
\end{figure}

Our final goal is to compare $Z(M_\ast)$ obtained from SFHs (i.e., mass and metallicities obtained from CMD fitting) with that obtained at higher redshift (i.e., metallicities from gas or integrated spectroscopy). However, to prevent any bias in obtaining the mass-metallicity relation from SFHs, it is fair to compare it first with $Z(M_\ast)$ obtained from local galaxies using spectroscopic metallicities of individual stars. To do so, we defined $M_\ast(t)$ and $Z(t)$  as the mass of the alive stars (that is, stars that at time $t$ have not still evolved into SNe or white dwarf phases) and the metallicity as a function of time, that is, the metallicity of the most recently stars formed at time $t$. We also define $\rm [Fe/H]_{RGB}$ as the mean metallicity of RGB stars in each galaxy. $M_\ast(t)$ and $Z(t)$ were obtained directly from the SFH for 14 fixed-time snapshots: $t$ = 1 to 13 Gyr in steps of 1 Gyr and at t=13.5 Gyr. The selection of the snapshot times was made according to the mean age resolution for $\psi(t)$ \citep[see][]{hidalgoetal2011}. $Z_{\rm RGB}$ was obtained as follows (we refer to \citet{hidalgoetal2011} for a complete description of the method). After obtaining the SFH, a synthetic CMD was created using $\psi(t)$ and $Z(t)$. Then, $\rm [Fe/H]_{RGB}$ was calculated by averaging the metallicity of the synthetic RGB stars down to a magnitude $M_{F814W}=-2$ mag. 

Assuming that the stars are formed with the same metallicity of the sourroundig gas, we used $Z(t)$ to compare the results
with gas-phase metallicities from the literature. On the other hand, $\rm [Fe/H]_{RGB}$ was used to compare with spectroscopic metallicities from RGB stars from the literature.

From an observational point of view, the two metallicities may differ. RGB metallicities are obtained form resolved red giant stars in the upper region of the RGB and represent averaged values of the metallicities that are locked up in the stars older than $\sim 2$~Gyr, while gas-phase metallicities are obtained from HII regions, which represent the metallicity of the most recently formed stars. To ilustrate this, Fig. \ref{f00} shows the difference between the two metallicities as a function of time for the entire LCID sample. Overall, the difference between $Z(t)$ and $\rm [Fe/H]_{RGB}$ increases with time, being similar at $\sim 2$~Gyr, as expected. We note that the evolution of $\rm Z(t) - [Fe/H]_{RGB}$  with time depends on how the galaxies evolve. Galaxies with a fast metallicity evolution reach a larger difference in $\rm Z(t) - [Fe/H]_{RGB}$  in a shorter time (e.g., Cetus and Tucana) than other galaxies with a slower evolution (e.g., Leo-A). Negative values of $\rm Z(t) - [Fe/H]_{RGB}$ would show stars that have been formed from fresh pre-enriched gas instead of the enriched gas present in the galaxy.

\begin{figure}
\centering
\includegraphics[width=8.4cm]{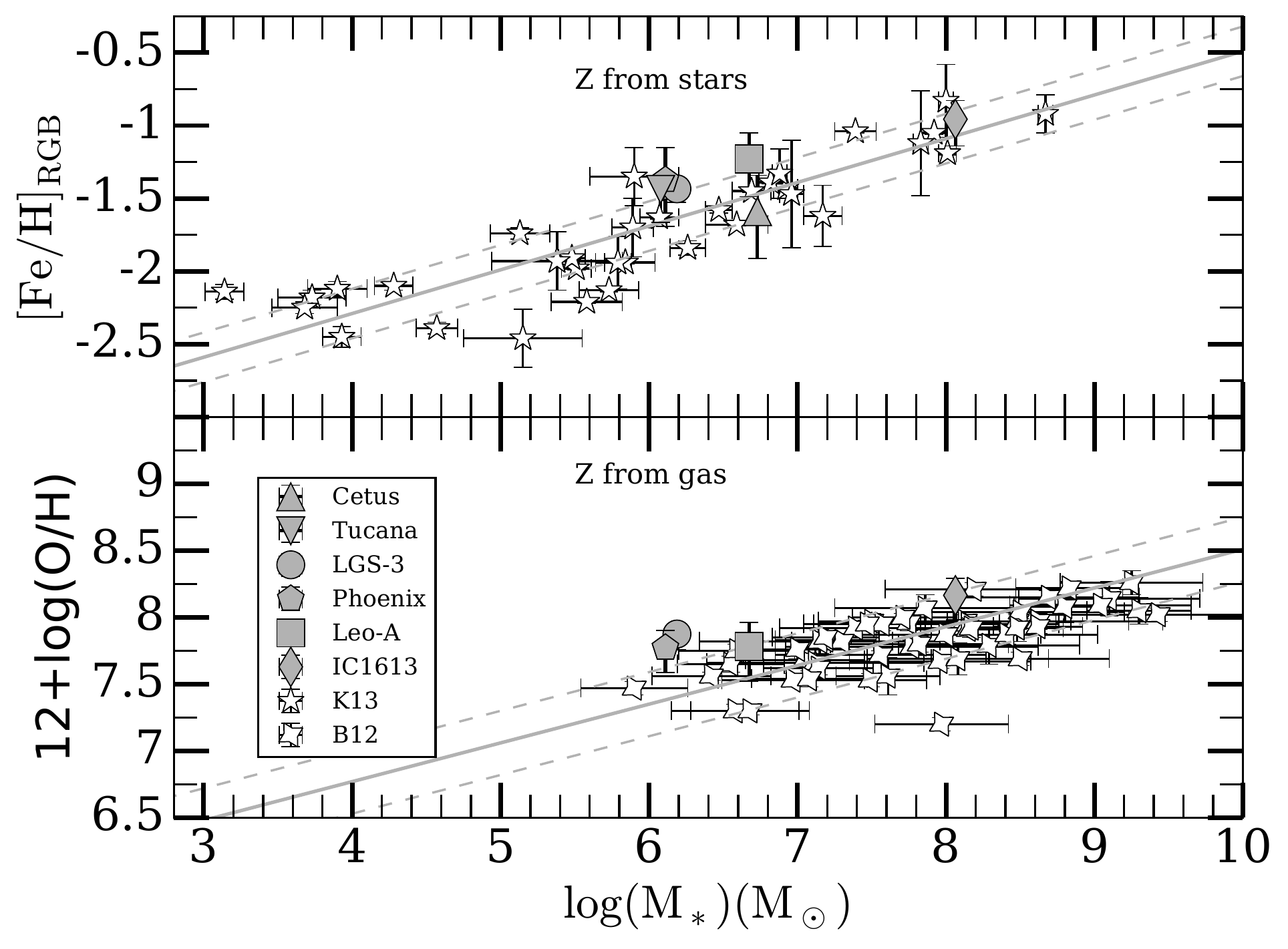}
\protect\caption[]{Mass-metallicity relation from spectroscopy of individual stars (upper panel) and from gas-phase metallicities (lower panel) compared with the results from the LCID dwarfs. The gray line shows the fit obtained by \citet{kirbyetal2013} (upper panel) and \citet{bergetal2012} (lower panel) in each case. For the LCID dwarfs, the $\rm [Fe/H]$ values have been converted into $\rm log(O/H) + 12$ using a solar scale.}
\label{f01}
\end{figure}

\begin{figure}
\centering
\includegraphics[width=8.4cm]{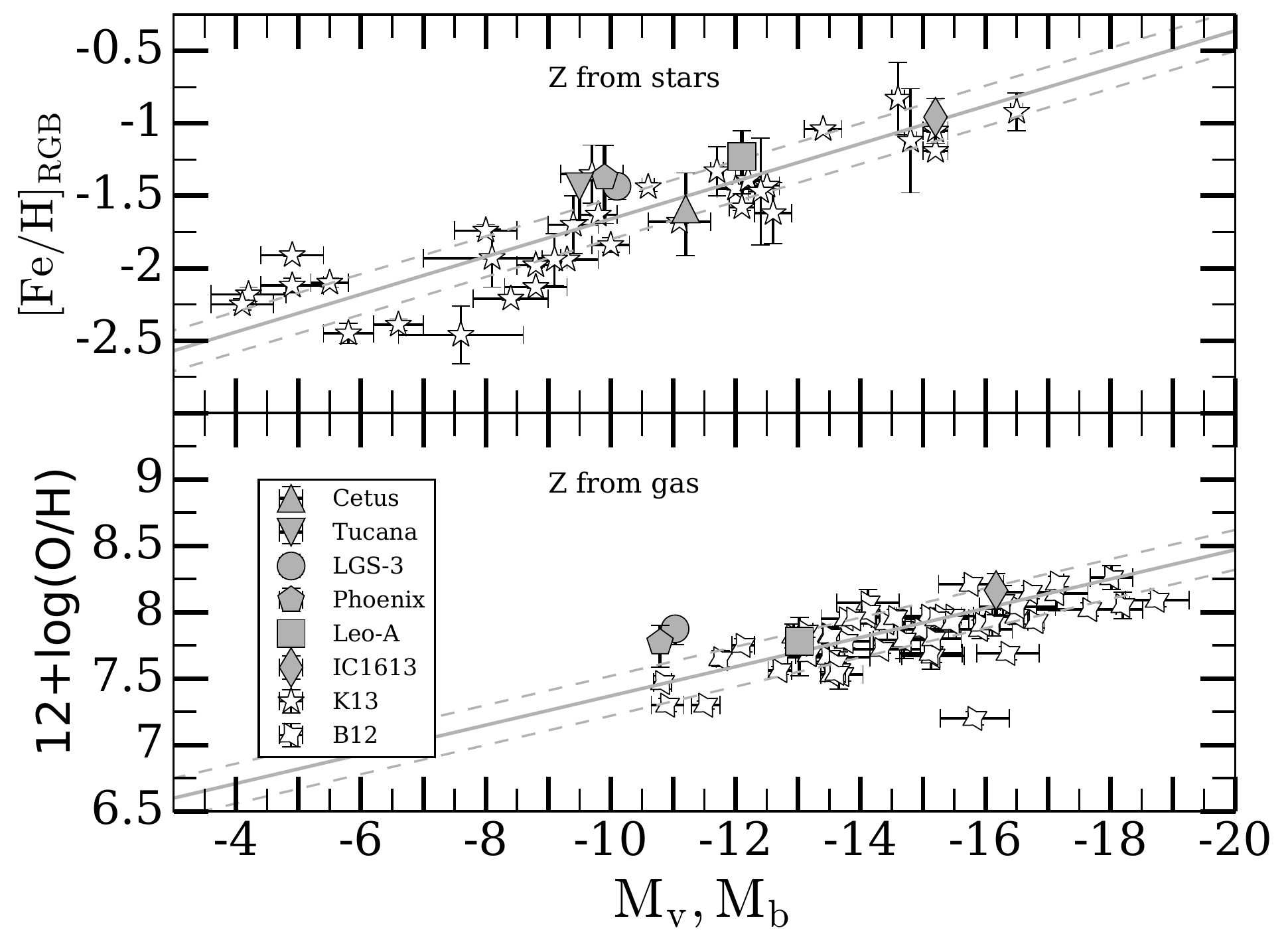}
\protect\caption[]{Luminosity-metallicity relation obtained from the LCID dwarfs compared with the results from \citet{kirbyetal2013} (upper panel) and \citet{bergetal2012} (lower panel). Only LCID dwarfs with non-negligible gas content are shown.}
\label{f02}
\end{figure}

\setcounter{table}{0}
\begin{table*}
\centering
\caption{Slope and intercept of $Z(M_\ast)$ from the literature.\label{t1}}
\begin{tabular}{lrrcccl}
\hline
$\rm Ref^{a}$      &Obs.                &Sample         &Redshift           &Zero-point           &$\alpha$               &$\rm M_B, M_\ast$ (range)                  \\
                   &                    &               &                   &                     &(KD02 calibration)    &(mag, $\rm log(M_\odot)$)                  \\\hline
BR12               &LVL + MMT           &31             &$0    \pm   0   $  &$5.24 \pm   0.60$    &$  0.33  \pm   0.07$   &$\rm M_\ast~(6.7~\textrm{--}~~\, 9.4)$     \\ 
CR12               &zCOSMOS             &6              &$0.25 \pm   0.05$  &$4.79 \pm   0.23$    &$  0.41  \pm   0.02$   &$\rm M_\ast~(8.6~\textrm{--}~ 9.6)$        \\     
CR12               &zCOSMOS             &5              &$0.63 \pm   0.05$  &$5.95 \pm   0.23$    &$  0.29  \pm   0.02$   &$\rm M_\ast~(9.2~\textrm{--}~ 10.5)$       \\  
EL08               &SDSS-DR4            &21             &$0.09 \pm   0.06$  &$5.72 \pm   0.14$    &$  0.33 \pm   0.02$    &$\rm M_\ast~(8.7~\textrm{--}~ 10.0)$       \\
ER06               &Keck I              &5              &$2.25 \pm   0.03$  &$6.13 \pm   0.17$    &$  0.25 \pm   0.02$    &$\rm M_\ast~(9.9~\textrm{--}~11.1)$        \\
POINT AND SLOPE
KB03               &DGSS                &18             &$0.48 \pm   0.04$  &$5.67 \pm   0.49$    &$ -0.15 \pm   0.02$    &$\rm M_B~(-17.6~\textrm{--} -21.3)$        \\
KB03               &DGSS                &18             &$0.66 \pm   0.03$  &$4.77 \pm   0.83$    &$ -0.20 \pm   0.04$    &$\rm M_B~(-18.5~\textrm{--} -21.7)$        \\
KB03               &DGSS                &9              &$0.78 \pm   0.03$  &$4.31 \pm   1.01$    &$ -0.21 \pm   0.05$    &$\rm M_B~(-19.0~\textrm{--} -22.6)$        \\
POINT AND SLOPE
KK04               &GOODS-N             &75             &$0.50 \pm   0.05$  &$3.75 \pm   0.56$    &$ -0.26 \pm   0.03$    &$\rm M_B~(-17.9~\textrm{--} -21.6)$        \\
KK04               &GOODS-N             &51             &$0.73 \pm   0.05$  &$2.79 \pm   1.30$    &$ -0.30 \pm   0.06$    &$\rm M_B~(-19.1~\textrm{--} -22.1)$        \\
KK04               &GOODS-N             &34             &$0.85 \pm   0.03$  &$2.36 \pm   0.86$    &$ -0.31 \pm   0.04$    &$\rm M_B~(-19.5~\textrm{--} -21.7)$        \\
LE06               &Spitzer + Lit.      &25             &$0    \pm   0   $  &$5.77 \pm   0.21$    &$ -0.14 \pm   0.01$    &$\rm M_B~(-13.3~\textrm{--} -21.1)$        \\
LE06               &Spitzer + Lit.      &25             &$0    \pm   0   $  &$5.53 \pm   0.16$    &$  0.31 \pm   0.02$    &$\rm M_\ast~(5.9~\textrm{--}~~\, 9.3)$     \\
PL01               &Lit.                &20             &$0    \pm   0   $  &$5.56 \pm   0.28$    &$ -0.16 \pm   0.02$    &$\rm M_B~(-11.2~\textrm{--} -15.5)$        \\
RM95               &MMT+CFHT            &24             &$0    \pm   0   $  &$5.56 \pm   0.48$    &$ -0.16 \pm   0.03$    &$\rm M_B~(-10.5~\textrm{--} -18.1)$        \\
SV05               &GDDS                &19             &$0.58 \pm   0.06$  &$5.31 \pm   0.75$    &$ 0.35 \pm   0.08$     &$\rm M_\ast(8.2~\textrm{--}~10.0)$         \\
SV05               &GDDS                &21             &$0.72 \pm   0.06$  &$3.78 \pm   0.77$    &$ 0.50 \pm   0.08$     &$\rm M_\ast(8.8~\textrm{--}~10.5)$         \\
SV05               &GDDS                &17             &$0.87 \pm   0.05$  &$3.94 \pm   0.77$    &$ 0.48 \pm   0.08$     &$\rm M_\ast(8.8~\textrm{--}~10.7)$         \\
SK89               &NOAO 2.1m           &19             &$0    \pm   0   $  &$5.48 \pm   0.27$    &$ -0.15 \pm   0.02$    &$\rm M_B~(-10.5~\textrm{--} -19.0)$        \\
TR04               &SDSS-DR2            &7              &$0.10 \pm   0.01$  &$6.06 \pm   0.19$    &$ 0.29 \pm   0.02$     &$\rm M_\ast~(8.5~\textrm{--}~~\, 9.2)$     \\
VZ97               &Kitt Peak 0.9m      &11             &$0    \pm      0$  &$5.05 \pm   0.40$    &$ -0.18 \pm   0.02$    &$\rm M_B~(-13.7~\textrm{--}~-18.5)$        \\
YT12               &SDSS-DR7            &89             &$0.08 \pm   0.03$  &$6.57 \pm   0.12$    &$ 0.25 \pm   0.01$     &$\rm M_\ast~(8.8~\textrm{--}~~\, 9.4)$     \\
ZH12               &SDSS-DR7            &30             &$0.07 \pm   0.02$  &$6.17 \pm   0.13$    &$ 0.28 \pm   0.01$     &$\rm M_\ast~(8.3~\textrm{--}~9.5)$         \\
ZH12               &DEEP2               &15             &$0.8  \pm   0.05$  &$5.87 \pm   0.18$    &$ 0.30 \pm   0.02$     &$\rm M_\ast~(9.2~\textrm{--}~~\, 10.6)$    \\\hline
\multicolumn{7}{l}{%
   \begin{minipage}{16cm}%
    \scriptsize [a] BR12:  Berg et al. (2012); CR12: Cresci et al. (2012); EL08: Ellison et al. (2008); ER06: Erb et al. (2006); KB03: Kobulnicky et al. (2003); KK04: Kobulnicky \& Kewley (2004); LE06: Lee et al. (2006); PL01: Pilyugin (2001); RM95: Richer \& McCall (1995); SV05: Savaglio et al. (2005); SK89: Skillman et al. (1989); TR04: Tremonti et al. (2004); VZ97: van Zee et al. (1997); YT12: Yates et al. (2012); ZH12: Zahid et al. (2012).%
   \end{minipage}%
}\\
\end{tabular}
\end{table*}

To compare the local mass-metallicity relation, $Z(M_\ast)$, with the relation obtained here from LCID dwarfs, we selected two papers with a large sample of local galaxies: \citet{kirbyetal2013}, who used spectroscopic metallicities obtained from individual RGB stars, and \citet{bergetal2012}, who used gas-phase metallicities. In order to minimize the scatter in the fit of the slope and intersect of the mass-metallicity relation for different works, we always used an ordinary least-squares bisector fit when the original data were available. In other cases, we used the values of the fit given by the authors.

Figure \ref{f01} (upper panel) shows $\rm [Fe/H]_{RGB}$ as a function of $M_\ast$ obtained from LCID dwarfs compared with the results from \citet{kirbyetal2013}. The LCID dwarfs fall within the $Z(M_\ast)$ obtained by \citet{kirbyetal2013}. The slope of $Z(M_\ast)$ computed using the LCID dwarfs is $\alpha=0.30\pm 0.06$, which is in full agreement with the slope computed by \citet{kirbyetal2013}, $\alpha=0.30\pm 0.02$. Two of the galaxies that are in both samples, IC1613 and Leo-A, show metallicities that are in agreement within the uncertainties (with differences of 0.23 dex and 0.35 dex, respectively). The $Z(M_\ast)$ obtained from \citet{bergetal2012} using gas-phase metallicities is also shown in Fig. \ref{f01} (lower panel) together with the $Z(t=0)$ obtained from LCID dwarfs, which falls within $Z(M_\ast)$. In this case, we removed Cetus and Tucana from the sample and only used the galaxies that show a non-negligible amount of gas and recent star formation. The mass-metallicity relation based on the LCID dwarf subsample is equal to $\alpha=0.20\pm 0.02$ for this case, slightly smaller, although still in fair agreement with the relation obtained by \citet{bergetal2012} ($0.29\pm 0.03$).

We also compared the luminosity-metallicity relation from the LCID dwarfs with the samples of \citet{kirbyetal2013} and \citet{bergetal2012}, who also reported $\rm M_v$ and $\rm M_b$ magnitudes, respectively. Figure \ref{f02} shows the luminosity-metallicity relation for both samples. The LCID dwarf samples also agree well in this case. This allows us to compare the luminosity-metallicity relation from the literature with the relation obtained here for the LCID dwarfs.

The agreement between metallicities obtained using SFHs with the metallicity obtained using spectroscopic metallicities of individual stars and gas-phase metallicities shows that we can use the results from LCID dwarfs to analyze the mass-metallicity relation. However, to analyze the dependency of the mass-metallicity relation on others parameters, metallicities obtained from gas are preferred since they reflect the physical events in the galaxies at the specific time period we consider, such as the SFR. We have shown that $Z(t=0)$ obtained from LCID dwarfs agrees well with the gas-phase metallicities shown in the literature, therefore we use $Z(t)$ in the next sections.

\section{MASS-METALLICTY RELATION AS A FUNCTION OF TIME}\label{secmmt}

To analyze the dependency of $Z(M_\ast)$ on redshift, we selected from the literature a large dataset providing reliable estimates of redshift, metallicities, magnitudes, and/or stellar masses. Only available data and well-established values were used. When necessary, all metallicities were transformed into the scale
of \citet[KD02, hereafter]{kewley&dopita2002}, as suggested by \citet{kewley&ellison2008}.

A least-squared bisector fit was performed to the available data to ensure the same type of fit for all publications. The fit of the slope of $Z(M_\ast)$, $\alpha$, was constrained to the range of stellar masses where $Z(M_\ast)$ is linear, avoiding the turn-over around $\rm log(M_\ast/M_\odot)=10.5$ where $Z(M_\ast)$ flattens \citep{yatesetal2012}. The same analysis was performed for the luminosity-metallitiy relation using published values of the absolute magnitude $M_B$ and metallicities.

Table \ref{t1} shows a summary of the selected publications. It gives the reference used (Col. 1), the name of the survey (or telescope) used in the publication (Col. 2), the number of galaxies used here for the fit (Col. 3), the mean redshift (Col. 4),  the intercept and the slope ($\alpha$) of the fit calculated by transforming the metallicity into the KD02 calibration (Cols. 5 and 6), and the range of the stellar masses or magnitudes used in the fit (Col. 7).

\begin{figure}
\centering
 \includegraphics[width=8.4cm,angle=0]{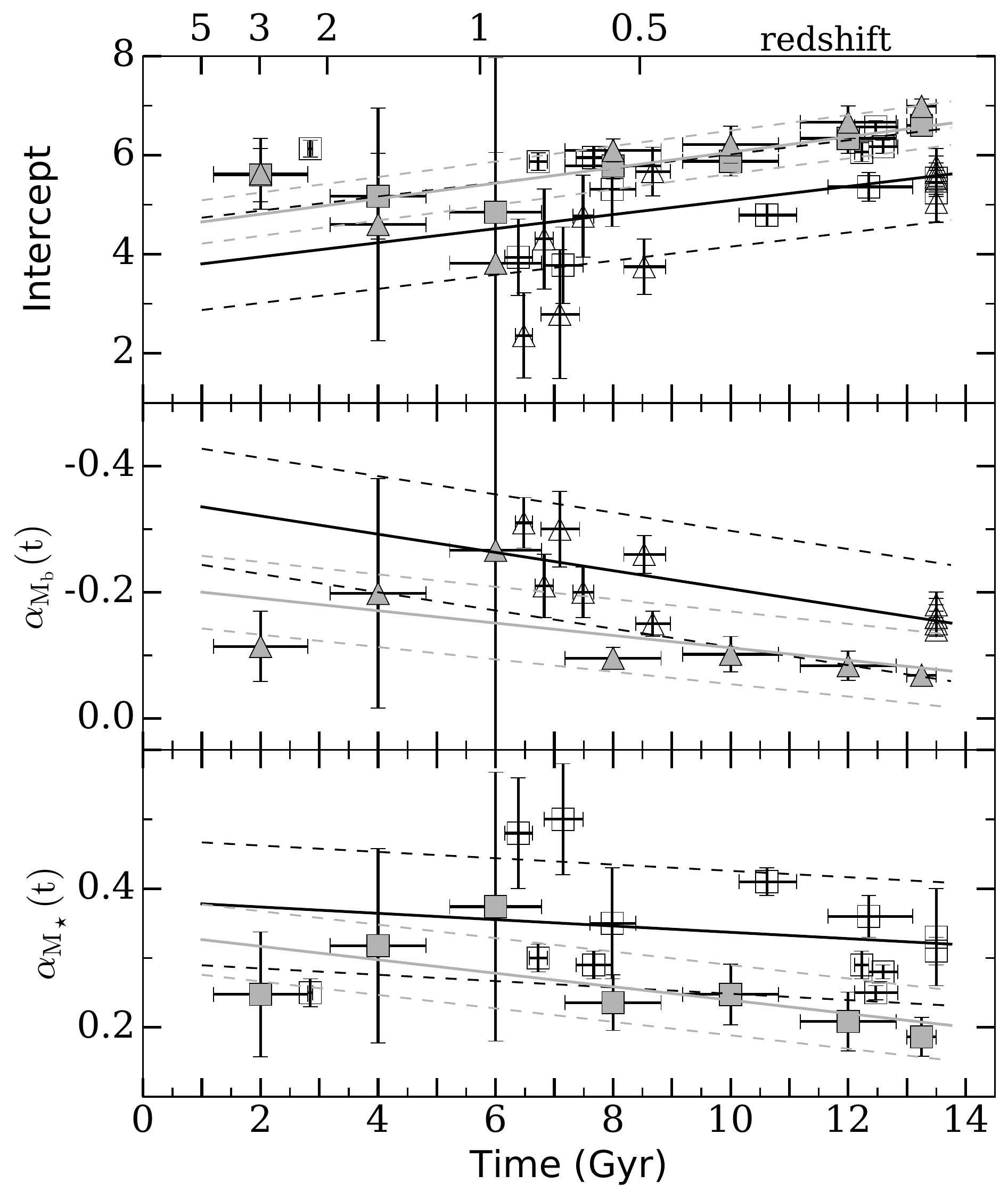}
\protect\caption[ ]{Intercept (upper panel), slope of the luminosity-metallicity (middle panel), and slope of the mass-metallicity relation (lower panel) as a function of time. The open symbols show results using gas-phase metallicities from literature. The filled symbols show data from LCID dwarfs. The data obtained from $M_B$ magnitudes are shown as triangles. The squares show results from stellar masses. The linear fits to the results are overplotted. The dashed lines show the uncertainties of the fit in each case. The vertical axis of the middle panel has been inverted to represent the same behavior as the others panels.}\label{f03}
\end{figure}

To compare our results with the values shown in Table \ref{t1}, we obtained the slope and intercept of the mass-metallicity and luminosity-metallicity from the SFHs of LCID dwarfs. To increase the statistics for the fit, we used time bins of 2 Gyr, which provide around 12 points per bin. Figure \ref{f03} shows the intercept (top panel) and $\alpha$ for the luminosity and mass-metallicity relation (middle and bottom panel, respectively) as a function of time for the data obtained from the literature and for the LCID dwarfs. 

To study the dependency on time of the variables shown in Fig. \ref{f03}, linear fits to the literature data and to the LCID dwarfs were obtained. The values of these fits are shown in Table \ref{t2}. Column 1 shows the data used for the fit (literature or LCID dwarfs data) for each one of the panels shown in Fig. \ref{f03}: intercept, luminosity-metallicity, and mass-metallicity relations. Column 2 shows the slope of the fit. Column 3 shows the Pearson correlation coefficient.

Both sets of fits show the same behavior: the slope becomes less steep, and the intercept increases with time. It is worthwhile noting that there is a larger dispersion of $\alpha$ obtained using stellar masses than the dispersion obtained using luminosities in the data from the literature, even after the converting the data into the same metallicity calibration. The luminosities are directly determined by observations, but stellar masses must be determined by assuming a mass-to-light ratio that in some cases may depend on the age of the dominant stellar population \citep{leeetal2006}. This may introduce a larger dispersion in the slope obtained using stellar masses compared to that obtained using luminosities. With these results, it is difficult to ensure a evolution of the mass-metallicity relation with time.

\begin{table}
\centering
\caption{Fits to the luminosity-metallicity and mass-metallicity relations shown in Fig. \ref{f03}.\label{t2}}
\begin{tabular}{lcc}
\hline
 Data                &Slope                            &$R^2$  \\\hline
 Intercept  (Lit.)   &$\,\,\,\, 0.142 \pm 0.078$       & 0.43  \\ 
 Intercept  (LCID)   &$\,\,\,\, 0.156 \pm 0.039$       & 0.71  \\
 $\rm M_b$  (Lit.)    &$\,\,\,\, 0.014 \pm 0.003$       & 0.75  \\
 $\rm M_b$  (LCID)    &$\,\,\,\, 0.010 \pm 0.005$       & 0.56  \\ 
 $\rm M_\ast$ (Lit.)  &$        -0.005 \pm 0.008$       & 0.19  \\
 $\rm M_\ast$ (LCID)  &$        -0.010 \pm 0.004$       & 0.62  \\\hline
\end{tabular}
\end{table}

\section{MULTIVARIATE MASS-METALLICTY RELATION}\label{secfmm}

There is tantalizing evidence of a dependency of $Z(M_\ast)$ on a second parameter such as the SFR \citep{yatesetal2012,mannuccietal2010,crescietal2012} or the HI mass \citep{bothwelletal2013, lara-lopezetal2013}, although this dependency might not be clear at high redshifts \citep{sandersetal2015}. If such a dependency exists, the dispersion of $Z(M_\ast)$ as a function of time shown in Fig. \ref{f03} could in part be due to an underlying dependency on additional parameters. The dependency of $Z(M_\ast)$ on additional parameters has been called the fundamental mass-metallicity relation by
various authors.

We performed a multivariate linear regression of the metallicity as a function of $M_\ast$, $\psi$, and $t$, that is, $Z(M_\ast,\psi,t)$. The values for $Z$, $M_\ast$, and $\psi$ were estimated using the SFHs of LCID dwarfs at the same time snapshots as defined in Sect. \ref{secmm}. Here we show the results for an exponential dependency of the form 
\begin{equation}\label{eq1}
           \rm Z = \alpha\, e^{a\,t}\,log(M_\ast)+\beta\, e^{b\,t}\,log(\psi) + K
\end{equation}
for different values of the exponents $a$ and $b$. We also analyzed a polynomial dependency on time, showing similar results for a second-order polynomial. However, the exponential function allows us to also analyze the case for no dependency on time by setting $a=0$ and $b=0$.

To obtain the coefficients $\alpha$ and $\beta$, we used a generalized least-squares model with a general covariance structure \citep{seaboldandperktold2010} in combination with a Monte Carlo simulation to obtain the exponents $a$ and $b$ that give the best results. 

Table \ref{t3} shows the result of the fit for six models. Column 1 shows the model number and Cols. 2 and 3 show the exponents $a$ and $b$, respectively. Columns 4 and 5 show the coefficients $\alpha$ and $\beta$, respectively. Column 6 shows the intercept. The Pearson correlation coefficient of the fit is shown in Col. 7. 
Models 1 to 3 show the case of no time dependency for $log(\psi)$ or $log(M_\ast)$ ($a=0$, $b=0$). Models 1 and 2 show only the
dependency on either $log(\psi)$ ($\beta=0$) or $log(M_\ast)$ ($\alpha=0$), respectively. Model 3 shows the dependency on both variables. Models 4 to 6 show a time dependency of the coefficients either for $log(\psi)$ (Model 4), $log(M_\ast)$ (Model 5), or both (Model 6). 

There is a clear improvement in the fit when a dependecy on time is introduced, which significantly increases the variability explained by the model from $\sim 60\%$ to $\sim 80\%$. In all the cases, the best fit for any model is obtained with $\beta < 0$, that is, for a fixed mass, galaxies with larger $\psi$ have lower metallicities. This effect has been observed by \citet{yatesetal2012} and \citet{crescietal2012}, for example, for low-mass galaxies.

\begin{figure}
\centering
 \includegraphics[width=8.4cm,angle=0]{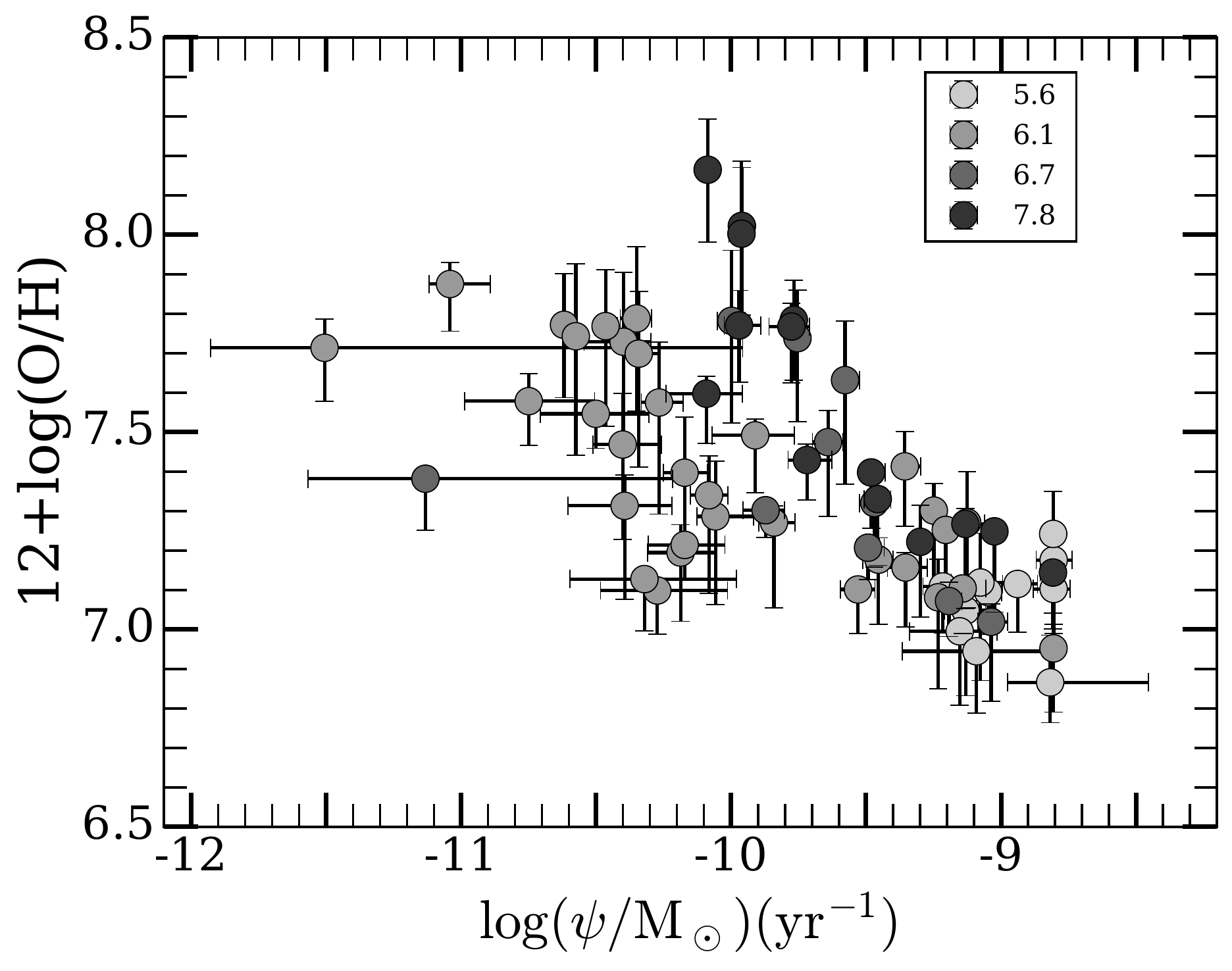}
\protect\caption[ ]{Metallicity as a function of the specific SFR. The colors represent different ranges of stellar masses.}\label{f04}
\end{figure}

To illustrate the contribution of $\psi$ to the metallicity for a fixed stellar mass, we represented the metallicity as a funcion of the SFR normalized to the stellar mass, that is, the specific SFR (sSFR), in Fig. \ref{f04}. For a selection of four stellar mass ranges of the LCID sample, it shows that galaxies with higher sSFR have lower metallicites.

\begin{table*}
\centering
\caption{Multivariate linear regression of the metallicity.\label{t3}}
\begin{tabular}{cllcccc}
\hline
Model & a           & \,\,b     & $\alpha$           & $\beta$              & K                  &$R^2$ \\\hline
  1   & 0           & \,\,0     &  fixed = 0         & $-0.92 \pm 0.36$     & $4.41 \pm 2.10$    &0.03  \\
  2   & 0           & \,\,0     & $ 0.45 \pm 0.05$   & fixed = 0            & $4.44 \pm 0.05$    &0.49  \\
  3   & 0           & \,\,0     & $ 0.44 \pm 0.05$   & $-0.27 \pm 0.04$     & $3.62 \pm 0.40$    &0.59  \\
  4   & 0           & \,\,0.141 & $ 0.23 \pm 0.02$   & $-0.03 \pm 0.01$     & $5.57 \pm 0.16$    &0.77  \\
  5   & 0.041       & \,\,0     & $ 0.14 \pm 0.01$   & $-0.08 \pm 0.02$     & $5.91 \pm 0.10$    &0.82  \\
  6   & 0.052       &-0.106     & $ 0.16 \pm 0.01$   & $-0.25 \pm 0.05$     & $5.41 \pm 0.19$    &0.85  \\\hline
\multicolumn{7}{l}{%
   \begin{minipage}{12cm}%
    \tiny Coefficients in $Z = \alpha\,e^{a\,t}\times log(M_\ast)+ \beta\,e^{b\,t}\times log(\psi) + K$ , with $t$ in (Gyr), $M_\ast$ in $(M_\odot)$, and $\psi$ in $(M_\odot~yr^{-1})$.
   \end{minipage}%
}\\
\end{tabular}
\end{table*}
 
\begin{figure}
\centering
 \includegraphics[width=8.4cm,angle=0]{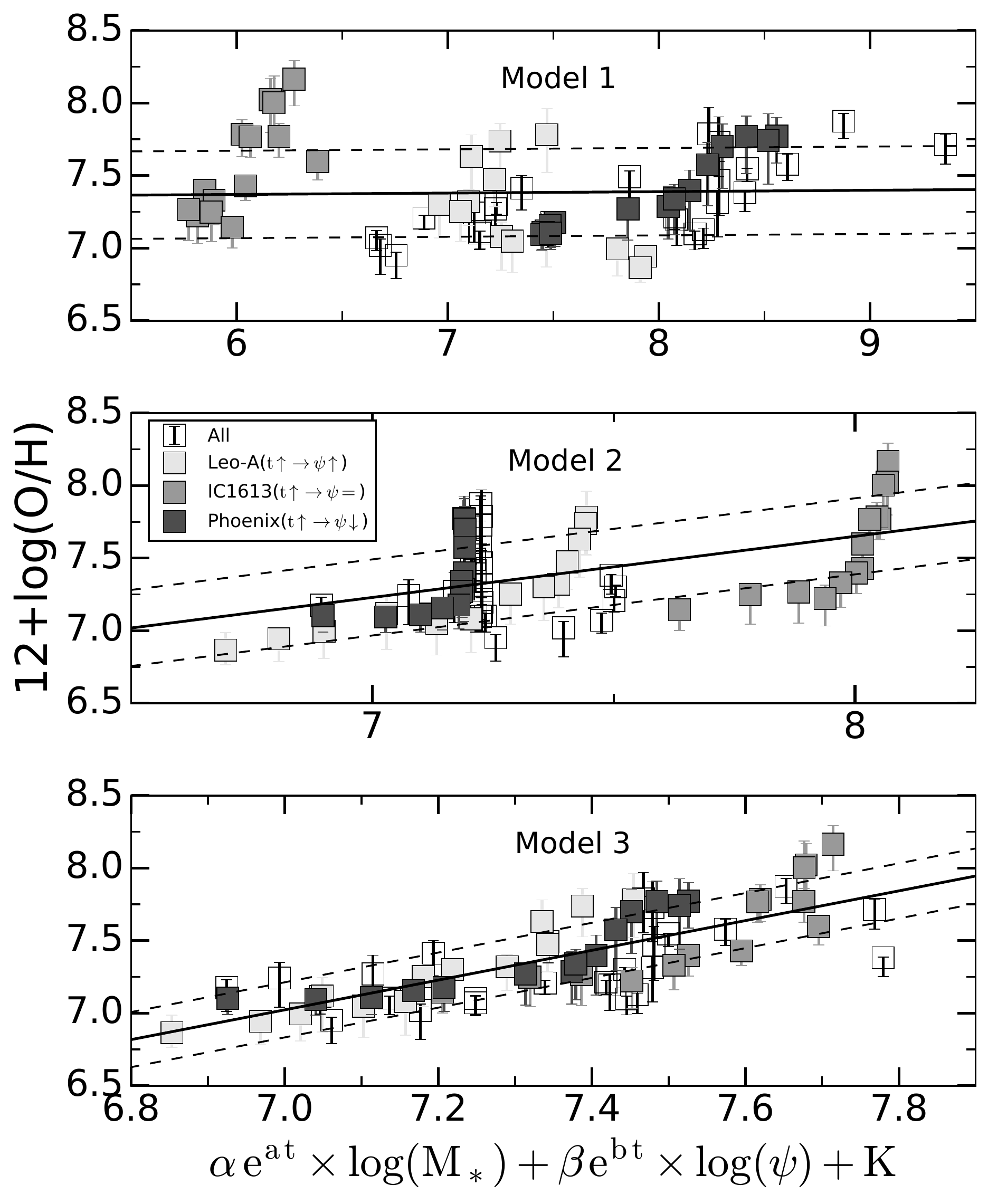}
\protect\caption[ ]{Projection of $Z = \alpha\,e^{a\,t}\times log(M_\ast)+ \beta\,e^{b\,t}\times log(\psi)$ for models 1 to 3 shown in Table \ref{t3}. Dashed lines show the uncertainties of the fit in each case. Three galaxies are highlighted, showing three different evolutions of $\psi$ with time: Leo-A, IC1613, and Phoenix (see text).}\label{f05}
\end{figure}

\begin{figure}
\centering
 \includegraphics[width=8.4cm,angle=0]{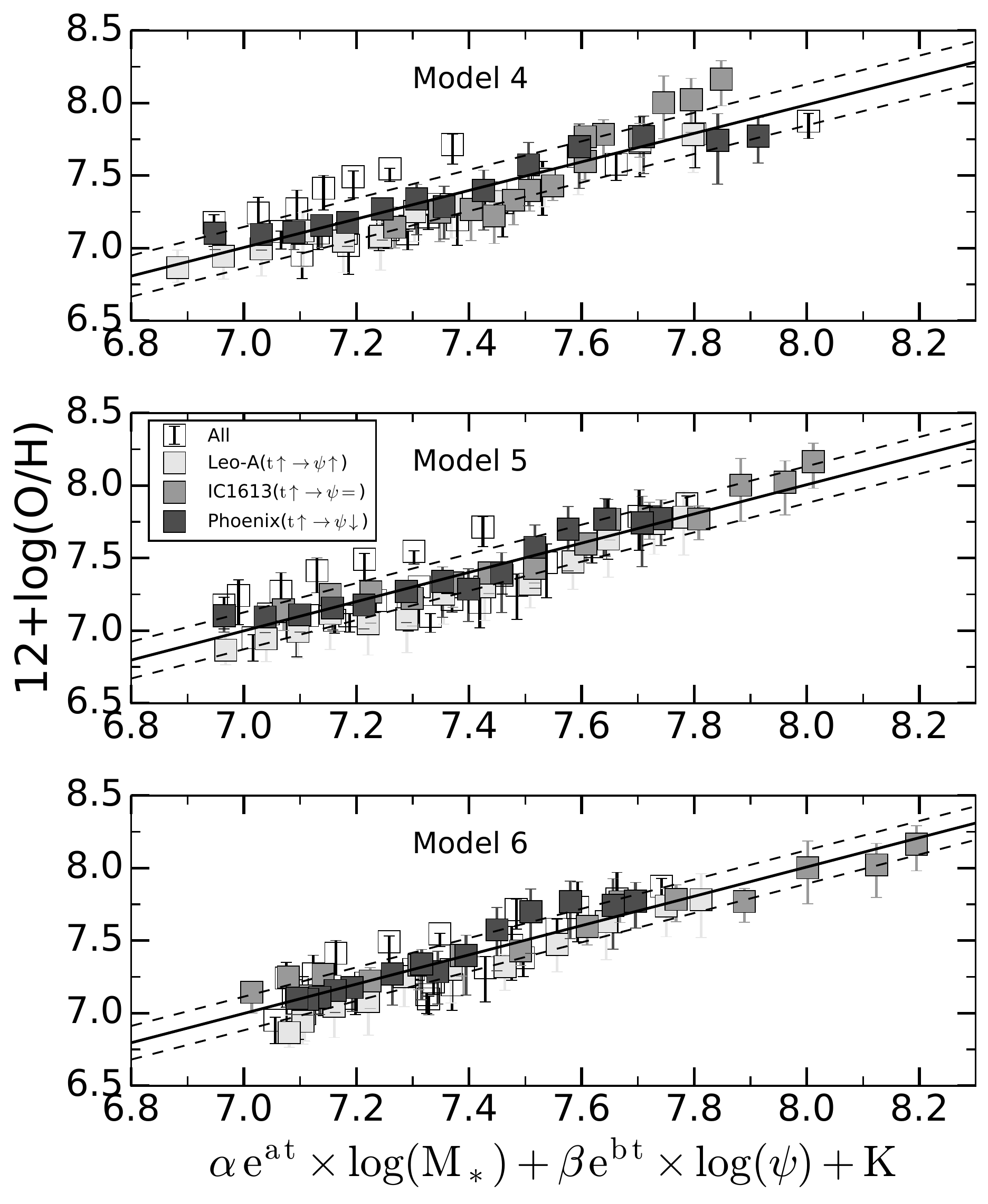}
\protect\caption[ ]{Same as in Fig. \ref{f05}, but for models 4 to 6.}\label{f06}
\end{figure}

Figures \ref{f05} and \ref{f06} show the projection of Eq. \ref{eq1} for all the models analyzed in Table \ref{t3}. The improvement of the fit is clearly shown when a dependecy on time is included; models 5 and 6 show the best results. Three galaxies are highlighted, Leo-A, IC1613, and Phoenix. Despite their completely different $\psi(t)$, the fit is good in the three cases.

\section{DISCUSSION AND CONCLUSIONS}\label{secdisc}

The dependency of the mass-metallicity relation on additional terms has been discussed by \citet{crescietal2012}, \citet{mannuccietal2010}, and \citet{yatesetal2012}, among others. These authors indicated a more general relation involving the SFR, that is, the so-called fundamental mass-metallicity relation. The relationship between $Z$, $M_\ast$, and $\psi$ obtained by these authors is very similar to the relation obtained here: $Z$ increases with $M_\ast$ ,
but decreases with $\psi$ for a fixed mass. 

On the other hand, \citet{bothwelletal2013} and \citet{lara-lopezetal2013} reported a dependency of the mass-metallicity relation also on the HI content of the galaxies in the sense that galaxies with a greater HI content have lower metallicities. \citet{lara-lopezetal2013} have noted that galaxies with a greater HI content are actively forming stars, which means that $\psi$ must be higher in these galaxies than in others with a low HI content. Therefore the results shown in these papers also agree well with the result obtained here.

The discrepancy between these results shown in the aforementioned papers could be explained as due to the different metallicity indexes and calibrations \citep{fosteretal2012}, the uncertainties associated to the estimation of the galaxy masses, or even the uncertainties in the estimation of the redshift. Transforming the original metallicity calibrators to a common calibration (see Table \ref{t1}) does not significantly change the dispersion in $\alpha$ with time found in the literature. The multivariate analysis performed in Sect. \ref{secfmm} shows that introducing a dependency on $\psi$  slightly increases the variability explained by the model (see Table \ref{t3}, models 2 and 3). The correlation coefficient increases significantly when a dependency of $\alpha$ on time is introduced (model 5). Little improvement is found when a dependency of $\beta$ on time is added to model 5 (i.e., model 6).

\citet{mannuccietal2010} and \citet{crescietal2012} reported that the observed evolution of the mass-metallicity relation with redshift is due to a selection biases at high redshift as well as to the increase in average SFR. However, the results shown here in Fig. \ref{f03} indicate that the evolution of the mass-metallicity relation seems to be unrelated to a change in $\psi$ with time, since three galaxies with different $\psi$ evolution show the same $Z(M_\ast,\psi,t)$ relation. This suggests that it does not play an important role for the type of the galaxies in the LCID sample how they are built in the $Z(M_\ast,\psi,t)$ relation.

It is beyond the scope of this paper to discuss the origin of the mass-metallicity relation or the implications of its evolution with time. Chemical evolution models would be necessary to interpret the age-metallicity relation of the LCID dwarfs shown here. Moreover, the number of LCID dwarfs analyzed here is not enough to derive general conclusions involving more massive galaxies. However, the accurate SFHs of the LCID dwarfs allow us to minimize the internal dispersion in the stellar mass, SFR, and metallicity estimations as a function of time, offering a new method to analyze the dependency of the mass-metallicity relation with other parameters. The results show an increasing contribution with time of the stellar mass of the galaxy to the metallicity of the stars most recently formed. However, the contribution of the SFR to the metallicity is almost constant with time, but opposite to the galaxy stellar mass: for a fixed mass, the metallicity decreases with increasing SFRs. While certainly speculative, this may suggest that the interstellar medium from which the stars are formed is enriched increasingly with time, as the stars of the galaxy evolve, recycling their metals to the medium. However, the constant contribution of the SFR to the metallicity may suggest fresh low-metallicity infall gas, which is mixed with the enriched interstellar medium and triggers the star formation.

The main conclusions of this paper are summarized as follows.

\begin{itemize}
 \item The luminosity-metallicity and mass-metallicity relation obtained from CMD-fitting techniques fully agree with the relation obtained from spectroscopy of resolved stars in the local environment. 
 \item We used stellar masses and gas-phase metallicities from the literature transformed into a common calibration to obtain the slope and the intercept of the mass-metallicity relation as a function of time. The result shows no evolution of the slope of the mass-metallicity relation with time. The same result has been obtained for LCID dwarfs.
 \item A multivariate analysis of the metallicity as a function of the stellar mass, SFR, and time suggests that the dependency of the metallicity on the galaxy stellar mass increases with time, with a direct correlation. However, the contribution of the SFR to the metallicity is almost constant with time, but with a inverse correlation: for a fixed mass, galaxies with higher SFRs have lower metallicities.
 \item The former results, although certainly speculative, may suggest that at the first evolutionary stages, the contribution to the metallicity enrichment in theses galaxies is driven mainly by the low-metallicity infall gas. This leads to a weak correlation between the metallicity of the galaxies and their stellar masses. Later on, as the galaxies evolve with time, the metallicity enrichment is driven mainly by the stellar mass of the system, leading to a stronger correlation of the mass-metallicity relation.
\end{itemize}

\section*{Acknowledgments}

I warmly thank the anonymous referee, who helped to improve the presentation of this paper. I am very grateful to A. Aparicio, S. Cassisi, and M. Cervi\~{n}o for their useful comments. The author is funded by the IAC (grant P/309403).

\bibliographystyle{aa}
\bibliography{hidalgo_final.bib}

\end{document}